\documentclass[12pt,dvips]{article}

\usepackage{amsmath,amssymb,exscale}
\usepackage{array,multicol}
\usepackage{afterpage,float,flafter}
\usepackage{epsfig,rotating,pifont}
\usepackage{cite}
\@addtoreset{equation}{section} \makeatother

\def\beq{\begin{eqnarray}}
\def\eeq{\end{eqnarray}}

\def\lsim{\mathrel{\rlap{\lower3pt\hbox{\hskip0pt$\sim$}}
    \raise1pt\hbox{$<$}}}         
\def\gsim{\mathrel{\rlap{\lower4pt\hbox{\hskip1pt$\sim$}}
    \raise1pt\hbox{$>$}}}         

\title{
\vspace{0cm}
\huge{Black Hole Bound on the Number of Species\\ and Quantum Gravity at LHC}
\vspace*{0.7cm}
\author{
\Large {\text{Gia Dvali$^{a,b}$} and  \text{Michele Redi$^{b,c}$}}\\ \\ 
\emph{$^a$ CERN, Theory Division, CH-1211 Geneva 23, Switzerland }\\
\emph{$^b$ CCPP, Department of Physics, New York University}\\
\emph{4 Washington Place, New York, NY 10003}\\
\emph{$^c$  ITPP, EPFL, CH-1015, Lausanne, Switzerland}}}

\date{}
\begin{document}
\maketitle \thispagestyle{empty} \vspace*{-.2cm}

\begin{abstract}

In theories with a large number $N$ of particle species, black hole
physics imposes an upper bound on the mass of the species equal to
$M_{Planck}/\sqrt{N}$. This bound suggests a novel solution to the
hierarchy problem in which there are $N \approx 10^{32}$
gravitationally coupled species, for example $10^{32} $ copies of
the Standard Model. The black hole bound forces them to be at the
weak scale, hence providing a stable hierarchy. We present various
arguments, that in such theories the effective gravitational cutoff
is reduced to $\Lambda_G \, \approx \,M_{Planck}/\sqrt{N}$ and a new
description is needed around this scale. In particular black-holes
smaller than $\Lambda_G^{-1}$ are already no longer semi-classical.
The nature of the completion is model dependent. One natural
possibility is that $\Lambda_G$ is the quantum gravity scale.
We provide evidence that within this type of scenarios, contrary to the standard intuition,
micro black holes have a (slowly-fading) memory of the species of origin.
Consequently the black holes produced at LHC, will predominantly
decay into the Standard Model particles, and negligibly into the other species.

\end{abstract}

\newpage
\renewcommand{\thepage}{\arabic{page}}
\setcounter{page}{1}

\section{Introduction}

  It is conceivable that the mysterious radiative stability
of the hierarchy between the Planck and the weak scales is imposed
by the intrinsic consistency of the underlying gravitational
dynamics. If so, this underlying principle should manifest itself as
a consistency relation between the different scales expressed in
terms of radiatively-insensitive quantities of the theory, such
as, topological numbers, representations of the symmetry groups, or
simply the number of species.

 An example of such a consistency bound, was proved recently in
\cite{bound}. It was shown that in theories with a large number $N$
of species with a mass scale $\Lambda$, black hole (BH) physics
imposes a relation between the Planck scale ($M_P$) and the mass of
the species ($\Lambda$)
\begin{equation}
\label{PM}
M_P^2 \, \gsim \, N \, \Lambda^2,
\end{equation}
up to a factor that scales as $\approx \log N$ at large $N$. In other
words, the large number of species weakens gravity by $1/N$.

\begin{figure}[ht]
\centering\leavevmode \epsfysize=6.5cm \epsfbox{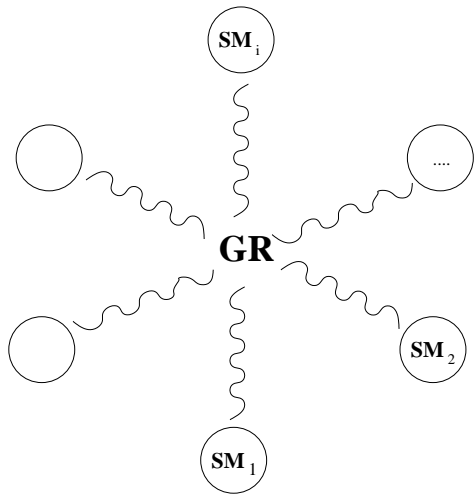}
\caption[Fig 1] {\it We consider as possible solution of the hierarchy
problem the existence of $10^{32}$ particle species interacting
gravitationally.}
\end{figure}

 Then, the  following simple solution to the hierarchy problem
emerges. Stability of the weak scale can be explained by the fact
that there are $N \approx 10^{32}$ species beyond the Standard Model
(SM). For instance, these new species could be $N$ copies of the
SM, perhaps related by permutation symmetry, and only
gravitationally coupled to each other.

  At first sight, from a perturbative point of view it might seem
mysterious what guarantees the stability of the mass of the particle
species. In the present work we provide various arguments that
the $N-$species effective theory breaks down at or below the reduced scale
\begin{equation}
\label{QG} \Lambda_G \approx M_P/\sqrt{N} \, .
\end{equation}
That is, when the bound (\ref{PM}) is saturated, $\Lambda_G$ approaches
the mass $\Lambda$ of the $N$ species and a completion is needed
right above that scale. From this point of view the fundamental scale of the
theory is $\Lambda_G$ while $M_P$ is a derived quantity whose large
value is due to the large multiplicity of species.

 One strong indication that new gravitational dynamics should appear
around the scale $\Lambda_G$ comes from the fact that the BHs of size
$\Lambda_G^{-1}$, much larger than Planck length, have lifetime
$\tau_{BH} \approx \Lambda_G^{-1}$ so they already probe the microscopic
theory of gravity. The reduced cutoff (\ref{QG}) in particular
reconciles the bound  (\ref{PM}) with naturalness arguments since in
absence of tuning the perturbative renormalization of the Planck
mass due to the $N$ species with cutoff $\Lambda_G$ saturates
(\ref{QG}). This idea was used in \cite{dgp2} as a possible origin of the
large value of $M_P$. This also agrees with perturbative arguments by
Veneziano \cite{veneziano}.

 The nature of the cutoff $\Lambda_G$ is not uniquely determined by
the consistency of the low energy theory. One natural possibility
that we investigate, is that gravity becomes strong and $\Lambda_G$ is the
quantum gravity scale. This is in fact the case in the Large Extra
Dimension scenario \cite{ADD} which can be understood as a special
case of our large $N$ theories. We focus on single-cutoff scenarios,
in which $\Lambda_G$ sets a universal cutoff both for
gravity as  well as for the other particle species (SM and its
copies). The most spectacular signature of this type of models is
that SM collisions at energies of order TeV will produce micro BHs
and the LHC will directly probe the quantum gravity regime. The
micro BHs will have very peculiar features. As we will show,
contrary to the naive intuition, BHs produced in particle
collisions at energy $\Lambda_G$ have a slowly-fading-away memory of
the particular copy of the species that produced them. As a result
the BHs produced at the LHC in the collision of SM particles will
predominantly decay into our SM particles and only with a
strongly suppressed probability into the other copies.

 We propose a dynamical interpretation to this fading-away-memory
in terms of slowly decaying quantum or classical hair of the
black holes. This hair is not associated with any long range
classical field. It either could be a hair under some discrete
symmetry, or under some massive gauge fields that are associated
with the given copy of  species. This slowly decaying BH hair then
creates a slowly fading-away memory of the site of
origin in the space of species. This implies that there is some notion of
locality in the space of species, although the only perturbative
link between the copies is the universally coupled graviton, and the
label of the copies  is not necessarily a coordinate in any real
space.

This paper is organized as follows. In section 2 we review the
argument determining the bound on the Planck scale. In section 3 we
discuss perturbative and non-perturbative arguments showing that
some new gravitational dynamics must appear below the scale
$\Lambda_G$. Section 4 contains examples of theories where the bound
on $M_P$ is saturated. Finally in section 5 we consider
the special properties of black holes in the theories with quantum
gravity scale $\Lambda_G$.

\section{The Black Hole Bound on the Number of Species}

 We shall first briefly reproduce the main idea behind the proof of
the bound of \cite{bound}.

The proof is the simplest when each sector of species has an
exactly conserved quantum number. Consider $N$ species of the
quantum fields $\Phi_j, ~~j= 1,2, ...N$, of mass $\Lambda$, each
carrying a separately conserved $Z_2$-charge. The system is
invariant under an exact discrete  $Z_2^N\, \equiv\, Z_2^{(1)}
\times Z_2^{(2)} \times \, ... \,Z_2^{(N)}$  symmetry, under the
independent sign flips of the fields.

In order to prove the relation (\ref{PM}) we can perform
the following thought experiment. Let us prepare a macroscopic BH
carrying the maximal $N$-units of the discrete charge. This can be
obtained by taking a large neutral BH and throwing one particle from
each species into it. The BH prepared in such a way will carry
exactly $N$-units of the conserved discrete charge. To avoid
entering the discussions on the black hole information loss issues,
it is useful to consider these $Z_2$-s as the gauged discrete
symmetries. As shown by Krauss and Wilczek \cite{ZN}, the
information about the absorbed charge then can be monitored by the
Aharonov-Bohm effect at infinity, using for example $Z_2$-cosmic
strings and cannot be lost.

 Because of the charge conservation, the information about the
$Z_2^N$-charge hosted by the black hole, must be revealed after its
evaporation. For a black hole with Hawking temperature $T_H$, the
probability of the emission of a heavy particle of mass $\Lambda \,
\gg \, T_H$ is exponentially suppressed by a Boltzmann factor $\approx
{\rm e}^{-{\Lambda \over T_H}}$. Thus, our black hole with $N$ units
of the $Z_2^N$-charge, can start emitting particles from the $N$ species,
only after its temperature drops to $T_H  \, \approx \, \Lambda$. At
this point, the mass of the black hole is $M_{BH}^* \, \approx  \,
{M_P^2 \over \Lambda}$. Starting from this moment, the black hole
can start revealing back the stored charge, in form of the $N$
particles species. However, by conservation of energy, the maximum
number of particles that can be emitted by the black hole is
\begin{equation}
\label{nmax}
n_{max} \, \approx \, {M_P^2 \over \Lambda^2}.
\end{equation}
These states should carry the same $Z_2^N$-charge  as the original
$N$-particles. Thus, $n_{max} \, = \, N$,  which proves the equation
(\ref{PM}).

In other words, the key point of the proof, is that the  amount of
the maximal discrete charge which is stored in the initial black
hole scales as $N$, but the temperature at which the black hole
starts giving back this charge essentially does not scale with
$N$. Hence the only way to avoid inconsistency is to scale the
Planck mass$^2$  as $N$. With minor assumption the same
conclusion holds generically in the presence of $N$ species.

\section{The Quantum Gravity Scale}

  The question that we wish to answer now is what is the cutoff
of the theory. We present several arguments that in the presence
of $N$ species new gravitational dynamics must appear below the
scale $\Lambda_G \approx M_P/\sqrt{N}$. While this holds in general, when
the BH bound (\ref{PM}) is saturated it implies that the theory must
break down right above the highest mass scale of the species.

\subsection{Black-Hole Argument}

 A strong non-perturbative argument supporting the claim that a
gravitational cutoff should be at the scale $\Lambda_G$ is provided
by the fact that the lifetime of the BHs of size $\Lambda_G^{-1}$ is
$\approx \Lambda_G^{-1}$.

 To see this, consider a BH with temperature $T_H \approx \Lambda_G$.
Let us assume first that the BH does not carry the conserved charge
of species number. In semi-classical approximation the decay rate of
the BH is,
\begin{equation}
\label{crate} {dM_{BH} \over dt} \, \approx \, -  \, N\,  T_H^2.
\end{equation}
Using the relation between the black hole mass and the temperature
$T_H \, \approx \, {M_P^2/M_{BH}}$, we can re-express this as,
\begin{equation}
\label{total} \tau_{BH} \, \approx \, {1 \over N}
\, \int_{0}^{M_{in}} \,  {M_{BH}^2 \over M_P^4}\, dM_{BH} \,
\end{equation}
where $M_{in}$ is the initial mass, which for the black hole of size
$\Lambda_G^{-1}$  is $M_{in} \, \approx  \, {M_P^2 \over \Lambda_G}$. This
gives,
\begin{equation}
\label{tbh} \tau_{BH} \, \approx  \, {M_{BH}^3 \over N M_P^4} \, \approx
\, {1\over \Lambda_G}.
\end{equation}

If the BH has $N$ units of the conserved charge, this calculation
has to be slightly modified to take into account the fact that due
to the conservation of charge the number of species available for
emission decreases as the BH evaporates. In this case,
\begin{equation}
\label{crate2} {dM_{BH} \over dt} \, \approx \, -  \, n(M_{BH}) \,
T_H^2,
\end{equation}
where  $n(M_{BH}) \, = \, {M_{BH} \over \Lambda_G} $ is the number of
available species. Going through the same steps as above one finds
that the lifetime of the BH is again given by (\ref{tbh}).

The fact that the lifetime becomes comparable to the size of the BH
implies that they cannot be treated as semi-classical four
dimensional BHs with well-defined Hawking temperature, since the
temperature $T_H$ itself changes on the time-scale $\approx
T_{H}^{-1}$. Therefore we conclude that $\Lambda_G$ determines the
critical scale beyond which new gravitational dynamics must appear.

\subsection{Perturbative Arguments}

We now turn to perturbative arguments (see \cite{veneziano} for
related work). To simplify the discussion let us assume that there
are $N$ species with mass $\Lambda \lsim \Lambda_G$ which are weakly
coupled. It is easy to see that the perturbation theory breaks down
at the scale $M_P/\sqrt{N}$.\footnote{We are grateful to Riccardo
Rattazzi for discussions about this point.} Consider the correction
to the graviton propagator due to the loops of the $N$ particles.
Neglecting the index structure this is given by,
\begin{equation}
\frac 1 {M_P^2} \frac 1 {p^2} <T(p) T(-p)> \frac 1 {p^2}
\end{equation}
where $T$ denotes the energy momentum tensor. The two point function
of $T$ has UV sensitive contribution and a non-local, calculable
piece associated to the production of the particles. The form of the
correction can be obtained by noting that well above the mass of the
particles the theory becomes approximately conformal. By conformal
invariance one has,
\begin{equation}
<T(p)T(-p)>= c\, p^4 \log \frac {p^2}{\Lambda^2}
\end{equation}
where the constant $c$ is the central charge counting
the number of degrees of freedom of the theory, i.e. $c \approx N$.
Comparing this with the tree level result one finds that
the perturbative expansion fails at the scale,
\begin{equation}
p^* \approx \frac {M_P}{\sqrt{N}} \label{pstar}
\end{equation}
This indicates that gravitational cutoff in a theory with $N$
species cannot be higher than $p^*$.

 This result reconciles the bound on the Planck mass with
naturalness arguments. The perturbative renormalization of the
Planck mass due to the species is given by the quadratically
divergent part of the two point function of $T$. In absence of other
contributions and assuming no cancelations, the net contribution
to the Planck mass is expected to be,
\begin{equation}
M_P^2\approx N \Lambda_G^2
\end{equation}
with $\Lambda_G$ the cutoff, which reproduces eq. (\ref{pstar}).

 Let us mention that the lowering of the cutoff
due to the large number of species resembles the situation in gauge
theories with positive $\beta-$function. Consider for example a
$U(1)$ gauge theory with $N$ flavor species. Above the
mass of the heaviest particle the one loop correction to the gauge
coupling is given by,
\begin{equation}
g(p)^2=g(\mu)^2+N  \frac {g(\mu)^4}{12 \pi^2}  \log \frac {p^2}
{\mu^2}
\end{equation}
which increases the gauge coupling at short distances. The
perturbative expansion goes out of control at the scale,
\begin{equation}
p^*= \mu\, e^{\frac {12 \pi^2}{g(\mu)^2 N}}
\end{equation}
This of course corresponds to the Landau pole of the theory below
which the effective theory must break down. In order for the
effective theory to have a minimal regime of validity clearly one
should have,
\begin{equation}
g(\mu)^2 N \lsim 12 \pi^2
\end{equation}
so that there is a separation between the Landau pole and $\mu$.
For a given $g(\mu)$ the cutoff gets reduced (exponentially) as we
increase the number of species. We should here note however that in
the gauge theory case with a large number of colors, the correction
has the opposite sign so that the coupling is reduced at short
distances and the corrections can be resummed.
It would interesting to see if there exists an analog in
the gravity case. A hint that this must be the case is given by the
fact that, while in string theory there exists an infinite tower of states,
nevertheless the graviton amplitudes are manifestly soft at high energy.
For ordinary matter with spin less or equal than one, it follows
from unitarity that the sign of the correction always makes gravity
stronger at short distances because the central charge must be
positive. One possibility is that this might be violated
in theories containing higher spins such as string theory. We leave
this questions for future work.

\subsection{Maximal Temperature}

 We would like now to show that the existence of the scale $\Lambda_G$
can also be inferred by considering the thermodynamics of the $N$ species.
The basic idea that we want to present is that a theory with $N$
species, has a limiting temperature set by $M_P/\sqrt{N}$,
or else it will be possible to prepare systems which violate
Bekenstein's entropy bound \cite{entropy}. The connection
between maximal temperature and entropy bounds was also
considered in \cite{veneziano2}. It would be also interesting to
explore the connection with other entropy bounds
\cite{otherbounds}.

\begin{figure}
\label{fig2} \centering\leavevmode \epsfysize=4cm \epsfbox{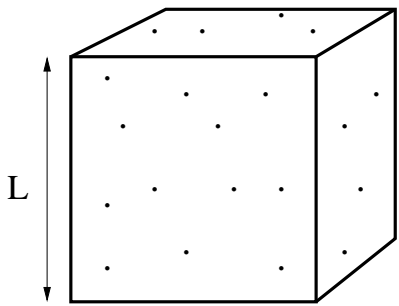}
\caption[Fig 2] {\it A box filled with radiation violates the entropy
bound prematurely.}
\end{figure}

Let us consider the following gedanken experiment, depicted in Fig. 2.
Take a box of size $L$ in which all the species are heated at temperature
$T$ much above their mass $\Lambda$ so that the particles can be treated
as relativistic. Under the assumption that thermal distribution for all
the species can be reached, the energy stored in the box is given
by\footnote{Clearly because of weak interspecies interactions,
particles may only thermalize within their own species or not
thermalize at all. However what is needed for the argument is that
they have a long enough maintained thermal distribution. This would
be the case for example if the particles are produced by
Gibbons-Hawking radiation in a time dependent background.},
\begin{equation}
E \approx N T^4 L^3.
\end{equation}
The size $L$ of the box is limited by the fact that the system will
eventually collapse into a black hole. Working in approximately
flat space, i.e. neglecting gravity back-reaction, is justified
if the Schwarzchild radius of the black hole with the same energy
is smaller than $L$. This gives the upper bound on the size of the
system,
\begin{equation}
\label{maxbox}
L_{MAX}\approx \frac {M_P}{T^2 \sqrt{N}}.
\end{equation}

We can now easily find a constraint on $T$. The entropy of the gas of
$N$ relativistic species is,
\begin{equation}
S_{gas}\approx N T^3 L^3.
\end{equation}
This entropy should not exceed the Bekenstein's bound \cite{entropy},
\begin{equation}
S \le S_{Bek}\approx M_P^2 L^2.
\end{equation}
By taking a box of maximal size (\ref{maxbox}) one finds that for this system there
exists a maximal temperature given by,
\begin{equation}
\label{tmax} T_{max}\lsim \frac {M_P}{\sqrt{N}}
\end{equation}
i.e. the particles cannot be heated above $\Lambda_G$.
In particular, if the hierarchy problem is explained by the large $N$
number of species $\Lambda_G\approx \Lambda$, so that it is impossible
to heat all the species at the temperature above their mass.
It is natural to interpret the existence of this maximal temperature
as a cut-off. In fact in theories with small number of species this
becomes simply the statement that $M_P$ is the limiting temperature.

 As an example let us consider the Large Extra Dimensions case \cite{ADD}.
In this context the previous bound can be understood as follow.
If we consider the higher dimensional theory with fundamental Planck
scale $M_D$ of order TeV this is obviously the maximum temperature
allowed for the system before compactification. In the experiment
considered above at high temperature the size of the box  (\ref{maxbox})
is much smaller that the size of the extra-dimension so that effectively we
are probing the theory in the decompactified limit. From the lower
dimensional point of view the limiting temperature is simply
understood as due to the collective effect of the tower of $4D$
Kaluza-Klein (KK) states. We would like to stress however that this bound
is derived in a model independent way and relies only on the
assumption that the system has a fixed number of degrees of freedom.

 It is interesting to note that an alternative derivation of the
maximal temperature can be obtained considering the following
simple cosmological setup\footnote{This approach relies on the assumption that the
standard $4D$ FRW cosmology is applicable up to the temperature of
interest. In particular  this is not the case in the Large
Extra Dimension scenario where gravity becomes higher dimensional at
much lower bulk temperature.}. Take an FRW universe filled with
$N$ species of mass $\Lambda$ or below whose temperature $\Lambda \,
\ll \, T \, \ll \, \Lambda_G$. Because the temperature is below the
quantum gravity scale, and species are relativistic, such Universe
will expand as radiation dominated FRW Universe with Hubble
parameter,
\begin{equation}
\label{Hubble}
H \, \approx \, {\sqrt{N}T^2 \over M_P}  \, \approx \, {T^2
\over \Lambda}.
\end{equation}

The first immediate upper bound on $T$ comes from the requirement
that Hubble be sub-Planckian, which gives  $T \, \approx  \,
\sqrt{M_P\Lambda}$. However a much more stringent bound arises from
noticing that the temperature of the relativistic fluid becomes
smaller than the horizon temperature $H$ at $T_{MAX}$ given by
eq. (\ref{tmax}). This should be clearly impossible since the quantum
fluctuations of the background will automatically increase the
temperature up to $H$.

\section{Examples}

 In this section we consider examples where the bound (\ref{PM}) is
saturated.

 First let us note that the black hole bound sheds a very
different light on the large extra-dimension solution to the
hierarchy problem. Here the weakness of gravity is
accounted by the large volume of the extra dimensional space, which
dilutes gravity and sets the hierarchy between $M_P$ and the
fundamental gravity scale $\Lambda_G$,
\begin{equation}
\label{add} M_P^2 \, = \, \Lambda_G^2  (\Lambda_G R)^n
\end{equation}
where $R$ is the typical size of the extra-dimensions, and $n$ is
their number. This relation acquires a very different meaning if we
note that the volume of the extra space determines the number of
KK species with mass $\Lambda_G$ in the following way,
\begin{equation}
(\Lambda_G R)^n \, = \, N.
\end{equation}
With this observation the relation
between the four dimensional and high dimensional Planck masses is
nothing but the saturation of the bound (\ref{PM}) (or equivalently
(\ref{QG}))! Looking from this point of view, ADD model solves the
hierarchy problem due to the fact that it includes $\approx 10^{32}$
KK gravitons of mass less than TeV.

 Hence, the Large Extra Dimension scenario can be regarded as a
particular example of a much larger class of models in which the hierarchy
problem is solved due to large number of species. Having $\approx 10^{32}$
copies of the SM is another example from this class, which has some advantages
with respect to higher dimensional geometric picture, since one does not have to
worry about the stabilization of the radius modulus. In addition the
cosmological constraints become less severe, due to the fact that,
unlike the production rate of KK gravitons, the production of other species
is very strongly suppressed.

 Another interesting example is provided by the Randall-Sundrum (RS)
scenario \cite{rs}. We focus for simplicity on the model with
one flat brane embedded in AdS$_5$. Although this model
does not solve the hierarchy problem due to the number of species,
nevertheless it saturates the bound on the Planck mass and the effective
number of degrees of freedom of the theory. From the $4D$ point of view
the model is described by a normalizable massless $4D$
graviton and a continuum of KK modes. The $4D$ Planck mass is given
by,
\begin{equation}
M_P^2= \frac {M_5^3} k \label{mprs}
\end{equation}
where $k$ is the AdS$_5$ curvature and $M_5$ the $5D$ Planck scale.
The physics of this model is greatly clarified by turning to the
holographic description. Using the AdS/CFT correspondence \cite{adscft},
RS is dual to a $4D$ strongly coupled ``large $N$'' conformal field theory
(CFT) coupled to gravity \cite{rscft}. The CFT is broken in the
ultraviolet at a scale $k$ and has a central charge related given
by $c=M_5^3/k^3$. While the CFT does not have particles in the
usual sense, given that the central charge measures the effective
number of degrees of freedom of the theory, one sees that
eq. (\ref{mprs}) can be written as,
\begin{equation}
M_P^2 \approx N k^2
\end{equation}
which saturates the bound (\ref{PM}) since the broken CFT has
excitations up to energy $k$.

It is important to clarify the nature of the cutoff in this
scenario. The description of RS in terms of a CFT coupled to gravity
is valid up to the scale $k$ where the CFT is broken. A possible UV
completion of the broken CFT is precisely given by the $5D$
geometric setup. At energies larger than $k$ the AdS curvature
becomes irrelevant and the physics turns higher dimensional. The
breakdown of the perturbative expansion is then simply understood as
the transition to the five dimensional regime. However in this case
gravity does not become strongly coupled till the fundamental Planck
scale $M_5$ which one can take arbitrarily larger than $k$.

More speculatively it is interesting to note that the bound between
the Planck scale and the number of species might be saturated in
perturbative string theory \cite{veneziano3}. Generalizing
the argument of \cite{bound} one finds that in arbitrary dimensions,
\begin{equation}
M_D^{D-2}\gsim N \Lambda^{D-2}.
\label{bhanyd}
\end{equation}
This is because the temperature of a BH of mass $M_{BH}$ in $D$ dimensions
is given by $T^{D-3}=M_D^{D-2}/M_{BH}$. From this it follows that the BH
can at most emit $M_D^{D-2}/\Lambda^{D-2}$ particles of mass $\Lambda$.
If there are $N$ species with approximately conserved species number to avoid inconsistencies
one finds that the Planck mass has to be scaled as in (\ref{bhanyd}).
Similarly the other arguments presented in Section 3 generalize to arbitrary
dimension by replacing $M_P^2$ with $M_D^{D-2}$ and scaling the other
dimensionfull quantities with the appropriate power.

Let us now consider critical string theory in $10D$. Here
one finds,
\begin{equation}
M_{10}^8 \approx \frac {m_s^8} {g^2}
\end{equation}
where $m_s$ is the string scale and $g$ the string coupling. It is
tempting to interpret $1/g^2$ as the effective number of degrees of
freedom so that the $M_{10}$ assumes the natural value. Note that this
is much less than the number of string modes between the $M_{10}$
and $m_s$ which grows exponentially. The reduction of number of
degrees of freedom has also been conjectured by Atick and Witten
\cite{hagedorn} in relation to the Hagedorn phase transition. The
existence of a limiting temperature $T_{Hag}\approx m_s$ in fact
also agrees with the temperature argument presented above.

\section{Strong Gravity at $\Lambda_G \approx TeV$}

 In the previous sections we have shown that in a theory with
$N$ species there exists an effective cutoff in the gravity sector
given by $\Lambda_G\approx M_P/\sqrt{N}$. One natural possibility
is that $\Lambda_G$ is the quantum gravity scale. We focus here on model
independent properties of black-holes in the scenarios with many species
solution to the hierarchy problem in which gravity becomes strong
at $\Lambda_G$.

 Since we assume that gravity becomes strong at the scale $\Lambda_G \, \approx \,
\Lambda$, one will start probing quantum gravity in high energy
particle collisions at the TeV scale. In particular, one should observe
a significant production of microscopic BHs at LHC. At first
sight this production leads us to a puzzle, the resolution
of which provides interesting informations about the effective BH
interactions in such theories.

  Consider a process with a characteristic energy $E$, in which a
particle-anti-particle pair of the $j-$th specie, is produced in the
annihilation of the two SM states (specie $i$), via the virtual BH exchange,
\begin{equation}
\label{process}
\psi_i \, + \, \bar {\psi_i} \, \rightarrow \,  BH  \rightarrow  \, \psi_j \, + \, \bar{\psi_j}\,.
\end{equation}
The rate of the process goes as,
\begin{equation}
\label{rate}
\Gamma_{i \rightarrow j} \, \approx \, {E^9 \over M_i^2 M_j^2 M_{BH}^4},
\end{equation}
where $M_i$ are the mass scales controlling the
effective BH coupling to the SM and to the other species.

Let us assume that $M_i$ are universally of order of
some typical scale $M$. To establish the bound on the scale $M$ we
consider the total production rate in which all possible
$j$ species are produced in the annihilation of the SM particles.
Assuming that the partial rates are $j-$independent, the
result is obtained by simple multiplication of (\ref{rate}) by $N
\approx M_P^2/\Lambda^2$
\begin{equation}
\label{ratetot}
\Gamma_{TOTAL} \, = \, \sum_j^N \, \Gamma_{i\rightarrow i} \, \approx \,
 {E^9 \over \Lambda^8} \left ({ M_P^2\Lambda^2 \over  M^4}\right) ,
\end{equation}
where we have used the fact that $M_{BH} \approx \Lambda$. Requiring
that the above rate does not violate unitarity at least until the
energy $E\approx \Lambda$, we get the bound
\begin{equation}
\label{bound1}
M \, \gsim  \, \sqrt{M_P\Lambda}
\end{equation}
In other words, if we assume universality of the couplings, unitarity implies
that the probability of production must be small in conflict with the
assumption of gravity becoming strong at $\Lambda$.

An even more severe lower bound on $M$ can be derived from the
following argument. Because the intermediate BH is heavy, below the scale
$\Lambda$ it can be integrated out. Then, the same diagram that was
generating the inter-species scattering process (\ref{process}),
generates an effective higher dimensional operator of the
form
\begin{equation}
\label{operator}
{T_{\mu\nu}^{(i)} \, T^{\mu\nu(j)}  \over M_i M_j M_{BH}^2}\, +\, ...
\end{equation}
where $T_{\mu\nu}^{(i)}$ is the energy momentum tensors of the $i-$th
specie. Other structures compatible with the symmetries, e.g.,
coupling between the vector or scaler currents, are of course also
possible.

 Since below the scale $\Lambda$, the above interaction can
be treated as a fundamental vertex, it will generate
quantum corrections to the Lagrangian of the SM fields. In
particular there will be a significant renormalization of the
kinetic terms of the SM fields from $\psi_j$ loops, due to their
enormous multiplicity. Putting aside accidental cancelations, the
wave-function renormalization factor of the SM fields will
scale as
\begin{equation}
\label{zeff}
Z_{eff} \, \gsim \frac {M_P^2} {M^2}.
\end{equation}
Requiring these effects to be small implies
\begin{equation}
\label{bound2}
M \,  \gsim \, M_P.
\end{equation}
Hence, the inter-species production rate through the BH exchange would be
as suppressed as the one mediated by graviton itself!

 The inevitable conclusion is that, either BHs should have suppressed
effective couplings to all the
species, or the assumption about the universality of this coupling
is wrong. The former option would be inconsistent with
gravity becoming strong at the scale $\Lambda$. Thus, we are lead to
the conclusion that BH coupling cannot be universal and somehow must
carry a label that allows them to differentiate among the species,
and predominantly decay into the specie of origin.

 Indeed, it is easy to see that the latter
possibility is compatible with the notion of
strong gravity at scale $\Lambda$ and makes the whole picture
consistent. Relaxing the assumption of universality of the scale
$M_j$, the total rate of production of species via BH exchange in
the collision of given $i$-th species  now becomes
\begin{equation}
\label{ratenew}
\Gamma_{TOTAL}^{(i)} \, = \, \sum_{j} \, \Gamma_{i\rightarrow j} \, \approx \,
 {1 \over M_i^2}{E^9 \over \Lambda^4} \left (\sum_j^{N} { 1 \over  M_j^2}\right) ,
\end{equation}
which puts a constraint
 \begin{equation}
\label{bound3} {1 \over M_i^2} \left (\sum_j^{N} { 1 \over
M_j^2}\right)  \, \lsim  \, {1 \over \Lambda^4}.
\end{equation}
Similarly the renormalization of the SM kinetic terms now gives the
bound
 \begin{equation}
\label{bound3}
 {1 \over M_i} \left (\sum_j^{N} { 1 \over  M_j}\right)  \, \lsim  \, \frac 1 {\Lambda^2}.
\end{equation}
Both bounds are automatically satisfied and reconciled with
gravity becoming strong at the scale $\Lambda$, if we assume that the BH
produced in the collision of $i$-th species decay to the same
species, with the rate $\approx \Lambda$, whereas the decay rate into
the other species is suppressed by $\approx \Lambda/M_P^2$.

  This means that micro BHs admit a certain notion of locality
in the space of species. The locality implies that BHs are endowed with a hair,
that gives them a memory of their site of origin in the space of species.
This hair may not necessarily be permanent, but it has to last long enough
in order to suppress the decay rate into other species. In such a case,
the hair will decay after certain characteristic time, and the memory about
the origin will fade away, allowing decay into foreign species.

  It is important to note that existence of such a hair is not in any
conflict with the standard classical BH no-hair theorems \cite{nohair}.
First, the hair that establishes the connection between the
BH and the site of origin can be quantum-mechanical and does not need to
be static. It was discovered by Krauss and Wilczek some time ago
\cite{ZN} that BH can carry a quantum mechanical hair under discrete gauge $Z_N$
symmetries. More recently, it was pointed out \cite{quantum}  that
in the presence of spin-2 or higher integer-spin massive fields, the
black holes may be endowed with the quantum hair under these fields.
Moreover, the BHs themselves are not classical BHs in the sense of
four dimensional gravity, since their gravitational radius is within their
Compton wavelength. Therefore, although the BHs
may appear classical from the point of view of an underlying
classical theory that operates at intermediate distances larger
than $\Lambda^{-1}$, they certainly are not classical in the usual sense.
They only become classical from the point of view of $4D$ gravity when
their mass reaches $M_P$, in which case universality of the decay is restored.

 To clarify the above statements we turn once more to the
Large Extra-Dimension scenario. There the relevant scales
are the radius of the extra dimensions $R$, the mass of a black hole $M_{BH}$, the
fundamental QG scale $\Lambda_G$ and the (four-dimensional) Planck
mass $M_P$. BHs of intermediate size $\Lambda_G \, \ll \,
M_{BH} \, \ll  \, M_P R^2$, have a size larger than $1/\Lambda_G$, but
smaller then the size of the extra dimension $R$. While such BHs can be
regarded as classical from the point of view of the high-dimensional
classical gravity operating and distances $\ll R$, in no sense
such objects are standard BHs from the point  of view of
a four-dimensional low energy observer, probing gravity
with the zero mode graviton. Therefore, neither four nor
higher dimensional observers would be surprised to discover that such
BHs are not universally coupled. And indeed, they are not. A micro
BH produced in a collision on a given brane, will mostly decay
into the species localized on the brane of origin and only with
very suppressed probability into states localized
at some $\approx R$ distance away. Only the very heavy BHs, with
masses $\gg M_P^2 R$, can reach across the bulk and start decay
universally. This of course agrees with the fact that these BHs
are classical already in the $4D$ description
and therefore universality is guaranteed by classical BH theorems.

 To conclude, while the simplest way to visualize locality is to imagine
the species separated in some physical space, what we are finding
is that locality in the space of species is something more general,
since at no point in our discussion we have assumed the existence
of underlying extra dimensions, in which the species are separated.

{\bf Acknowledgments}

We thank R. Brustein, G. Gabadadze, O. Nielsen, O. Pujolas, R. Rattazzi and G. Veneziano
for very useful discussions and comments. The work of G.D. and M.R. is supported in
part by David and Lucile Packard Foundation Fellowship for  Science
and Engineering, and by NSF grant PHY-0245068.



\begin{thebibliography}{99}
\bibitem{veneziano3}
We understand that this has also been conjectured by G. Veneziano.
\end{thebibliography}
\end{document}